\documentclass[preprint,12pt,authoryear]{elsarticle}

\usepackage{amssymb}
\usepackage{listings}

\journal{Frontiers in Neuroscience}

\usepackage{color}
\usepackage[normalem]{ulem}
\usepackage[export]{adjustbox}
\usepackage{subfig}
\usepackage{longtable}
\usepackage{float}
\usepackage[font=footnotesize]{caption}
\usepackage{url,lineno,microtype, eqnarray}
\usepackage[onehalfspacing]{setspace}

\usepackage{amsmath}

\usepackage{bookmark}

\begin{document}
\begin{frontmatter}

\title{A neuro-inspired system for online learning and recognition of parallel spike trains, based on spike latency and heterosynaptic STDP\\ {\normalsize preprint version}}

%INIZIO NOMI
\author[label1,label2]{Gianluca Susi}

\author[label1,label3]{Luis Ant\'on Toro} 

\author[label1,label4]{Leonides Canuet} 

\author[label1,label3]{Maria Eugenia L\'opez}  

\author[label1,label3]{Fernando Maest\'u} 

\author[label5]{Claudio R. Mirasso} 

\author[label1,label6]{Ernesto Pereda\corref{mycorrespondingauthor}}
\ead{eperdepa@ull.edu.es}

%FINE NOMI

%INIZIO CORRESPONDING
\cortext[mycorrespondingauthor]{Corresponding author: Ernesto Pereda}
%FINE CORRESPONDING

%INIZIO AFFILIAZIONI
\address[label1]{ UCM-UPM Laboratory of Cognitive and Computational Neuroscience, Center for Biomedical Technology (CTB), Technical University of Madrid, Spain}
\address[label2]{ Dipartimento di Ingegneria Civile e Ingegneria Informatica (DICII), Università di Roma `Tor Vergata', Italy;}
\address[label3]{Departamento de Psicología Experimental, Facultad de Psicología, Universidad Complutense de Madrid, Spain;}
\address[label4]{Departamento de Psicología Clinica, Psicobiología y Metodología, Universidad de La Laguna, Tenerife, Spain;}
\address[label5]{Instituto de Fisica Interdisciplinar y Sistemas Complejos, CSIC-UIB, Campus Universitat de les Illes Balears E-07122, Palma de Mallorca, Spain;}
\address[label6]{Departamento de Ingeniería Industrial, Escuela Superior de Ingeniería y Tecnología \& IUNE, Universidad de La Laguna, Tenerife, Spain}
%FINE AFFILIAZIONI

\begin{abstract}

Humans perform remarkably well in many cognitive tasks including pattern recognition. However, the neuronal mechanisms underlying this process are not well understood. Nevertheless, artificial neural networks, inspired in brain circuits, have been designed and used to tackle spatio-temporal pattern recognition tasks.

In this paper we present a multineuronal spike pattern detection structure able to autonomously implement online learning and recognition of parallel spike sequences (i.e., sequences of pulses belonging to different neurons/neural ensembles). 
The operating principle of this structure is based on two spiking/synaptic neurocomputational characteristics: \emph{spike latency}, that enables neurons to fire spikes with a certain delay and \emph{heterosynaptic plasticity}, that allows the own regulation of synaptic weights.
From the perspective of the information representation, the structure allows mapping a spatio-temporal stimulus into a multidimensional, temporal, feature space. In this space, the parameter coordinate and the time at which a neuron fires represent one specific feature. In this sense, each feature can be considered to span a single temporal axis.\\
We applied our proposed scheme to experimental data obtained from a motor-inhibitory cognitive task. 
The test exhibits good classification performance, indicating the adequateness of our approach. 
In addition to its effectiveness, its simplicity and low computational cost suggest a large scale implementation for real time recognition applications in several areas, such as brain computer interface, personal biometrics authentication or early detection of diseases.

\end{abstract}

\begin{keyword}
coincidence detection \sep spiking neurons\sep spike latency \sep delay \sep heterosinaptic plasticity \sep STDP \sep Go/NoGo
%% keywords here, in the form: keyword \sep keyword

%% PACS codes here, in the form: \PACS code \sep code

%% MSC codes here, in the form: \MSC code \sep code
%% or \MSC[2008] code \sep code (2000 is the default)

\end{keyword}

\end{frontmatter}

%% \linenumbers

%% main text
\section{Introduction}

In recent years there has been an increasing interest in applying artificial neural networks to solve pattern recognition tasks. 
However, it remains challenging to design more realistic spiking neuronal networks (SNNs) which use biologically plausible mechanisms to achieve these tasks \citep{Diehl2015}.
In sensory systems, the recognition of stimuli is possible by detecting spike patterns during the processing of peripheral inputs. 
Precise spike timings of neural activity have been observed in many brain regions, including the retina, the lateral geniculate nucleus, and the visual cortex, suggesting that the temporal structure of spike trains serves as an important component of the neuronal representation of the stimuli \citep{Gutig2006, Zhang2016}. 
Specific neural mechanisms that recognize time-varying stimuli by processing spiking activity have been an important subject of research \citep{Larson2010, Masquelier2017}. 
Whereas some investigations are oriented to the study of the spike activity of single neurons, many others consider the timing of spikes across a population of afferent neurons\citep{Gautrais1998, Stark2015} .

Plasticity regulates the strength in the connection between neurons.
In homosynaptic plasticity the activity in a particular neuron alters the efficacy of the synaptic connection with its target. 
Instead, in heterosynaptic plasticity changes in the synaptic strength can occur in both stimulated and non-stimulated pathways reaching the same target neuron.
Like homosynaptic plasticity, heterosynaptic plasticity has two forms: inhibition and potentiation \citep{Squire2013}; the latter not necessarily restricted to a subset of cells, but it can occur to many of the neurons in the population \citep{Han2013}. 
A number of distinct subtypes of heterosynaptic plasticity have been found in a variety of brain regions and organisms. 
They are associated to different neural processes including the development and refinement of neural circuits \citep{Vitureira2012}, extending the lifetime of memory traces during ongoing learning in neuronal networks \citep{Chistiakova2009}. 
Among these, heterosynaptic modulation (i.e., when the activity of a modulatory neuron induces a change in the synaptic efficacy between another neuron and the same target cell \citep{Phares2006}) allows that one set of inputs exert long-lasting heterosynaptic control over another, enabling the interplay of functionally and spatially distinct pathways \citep{Han2013}. 
Among the various types of heterosynaptic plasticity, the heterosynaptic form of Spike-Timing-Dependent Plasticity (STDP) is capturing a lot of interest because recent works have shown that it is a critical factor in the synaptic organization and resulting dendritic computation \citep{Hiratani2017}. \\
%Excitatory lateral connections, i.e., connections between neurons that belong to parallel paths, are though to be fundamental to link neurons with similar receptive fields properties \citep{Buzas2006} providing a potentially powerful mechanism to enhance coordinated activity \citep{Christie2006}.\\
In this paper we introduce a simple but effective network topology specialized in online recognition of temporal patterns. The structure is characterized by lateral excitation, i.e., excitatory connections between neurons that belong to parallel paths, and is based on two features: heterosynaptic STDP and \emph{spike latency}. 
Neurons dynamics is described using the Leaky Integrate-and-Fire with Latency (LIFL) model, which is similar to the Integrate and Fire but supports the spike latency mechanism, extracted from the more realistic Hodgkin-Huxley (HH) model \citep{Salerno2011}.
The structure maps spatio-temporal stimuli to specific areas in a temporal, multidimensional, feature space. 
In addition it is able to self-regulate its weights, allowing the learning and recognition of multineuronal temporal patterns in parallel spike trains arising from neuronal ensembles. 
% In this way, identified objects are described through attribute categories, and each category is topologically structured with elementary building blocks among repetitive cortical columns.\\
In order to show the potential of the presented structure, we apply it to a cognitive task-recognition problem, considering magnetoencephalografic (MEG) signals of subjects while performing a Go/NoGo task.

%For Original Research Articles \citep{conference}, Clinical Trial Articles \citep{article}, and Technology Reports \citep{patent}, the introduction should be succinct, with no subheadings \citep{book}. For Case Reports the Introduction should include symptoms at presentation \citep{chapter}, physical exams and lab results \citep{dataset}.

\section{Materials and Methods}

\subsection{LIFL neuron model}
\label{SECT:NeuMod}
The LIFL neuron model differs from the Leaky Integrate-and-Fire (LIF) model because it includes the \emph{spike latency} \citep{Izhikevich2004, Cristini2015, Susi2016} neurocomputational feature. 
It consists of a membrane potential-dependent delay time between the overcoming of the ``threshold'' and the actual spike generation \citep{Izhikevich2004, Salerno2011}. 
This delay is important because it allows encoding the strength of the input in the spike times \citep{Izhikevich2007} extending the neuron computation capabilities over the threshold \citep[e.g.,][]{Gollisch2008, Fontaine2009, Susi2015}. 
Neurons with such feature are present in many sensory systems, including the auditory, visual, and somatosensory system \citep{Wang2013, Trotta2013}. 
The LIFL neuron model embeds spike latency using a mechanism extracted from the more realistic Hodgkin-Huxley model \citep{Salerno2011}. 
It is characterized by the internal state $S$ (representing the membrane potential), that ranges, for simplicity, from $0$ (corresponding to the resting potential of the biological neuron) to $\infty$.

In its basic implementation, the LIFL model uses a defined threshold ($S_{th}$), a value slightly greater than $1$ that separates two different operating modes: a \emph{passive mode} when $S<S_{th}$, and an \emph{active mode} when $S>S_{th}$. 
In the passive mode, $S$ is affected by a decay, whereas in the active mode it is characterized by a spontaneous growth. 
For simplicity, we assume that simple Dirac delta functions (representing the action potentials) are exchanged between neurons, in form of pulses or pulse trains. 

The LIFL can be implemented through the event-driven technique \citep{Mattia2000}, that provides fast simulations \citep{Ros2006}. 
When the postsynaptic neuron $N_j$ receives a pulse from the presynaptic neuron $N_i$, its internal state is updated through one of the following equations, depending on whether $N_j$ is in the passive or in the active mode, as:
\begin{equation}
S_{_{Nj}}= S_{p\;_{Nj}}+A_{_{N_i}}\cdot w(N_j,N_i)-T_l \; , \mbox{ for } 0\leq S_{p\;_{Nj}}<S_{th} \label{EQ:Pm}
\end{equation}
\begin{equation}
S_{_{Nj}}= S_{p\;_{Nj}}+A_{_{N_i}}\cdot w(N_j,N_i)+T_r \; , \mbox{ for } S_{th}\leq S_{p\;_{Nj}}< S_{max} \label{EQ:Am}
\end{equation}
$S_{p\;_{Nj}}$ represents the postsynaptic neuron's \emph{previous state}, i.e., the internal state immediately before the new pulse arrives.
$A_{_{N_i}}$ represents the amplitude of the generated pulse; $w(N_j,N_i)$ represents the \emph{synaptic weight} from neuron $N_i$ to neuron $N_j$. 
The product $A_{_{N_i}}\cdot w(N_j,N_i)$ represents the amplitude of the post-synaptyc pulse arriving to $N_j$. 

$T_l$, the \emph{leakage term}, accounts for the underthreshold decay of $S$ during two consecutive input pulses in the passive mode. 
We will consider LIFL basic configuration, i.e., characterized by a linear subthreshold decay \cite[as in][]{Mattia2000}, where $T_l = L_d \cdot \Delta t$, being $L_d$ a non negative quantity called \emph{decay parameter} and $\Delta t$  the temporal distance between two consecutive incoming spikes.

\begin{figure}
\centering
\includegraphics[width=0.5\textwidth]{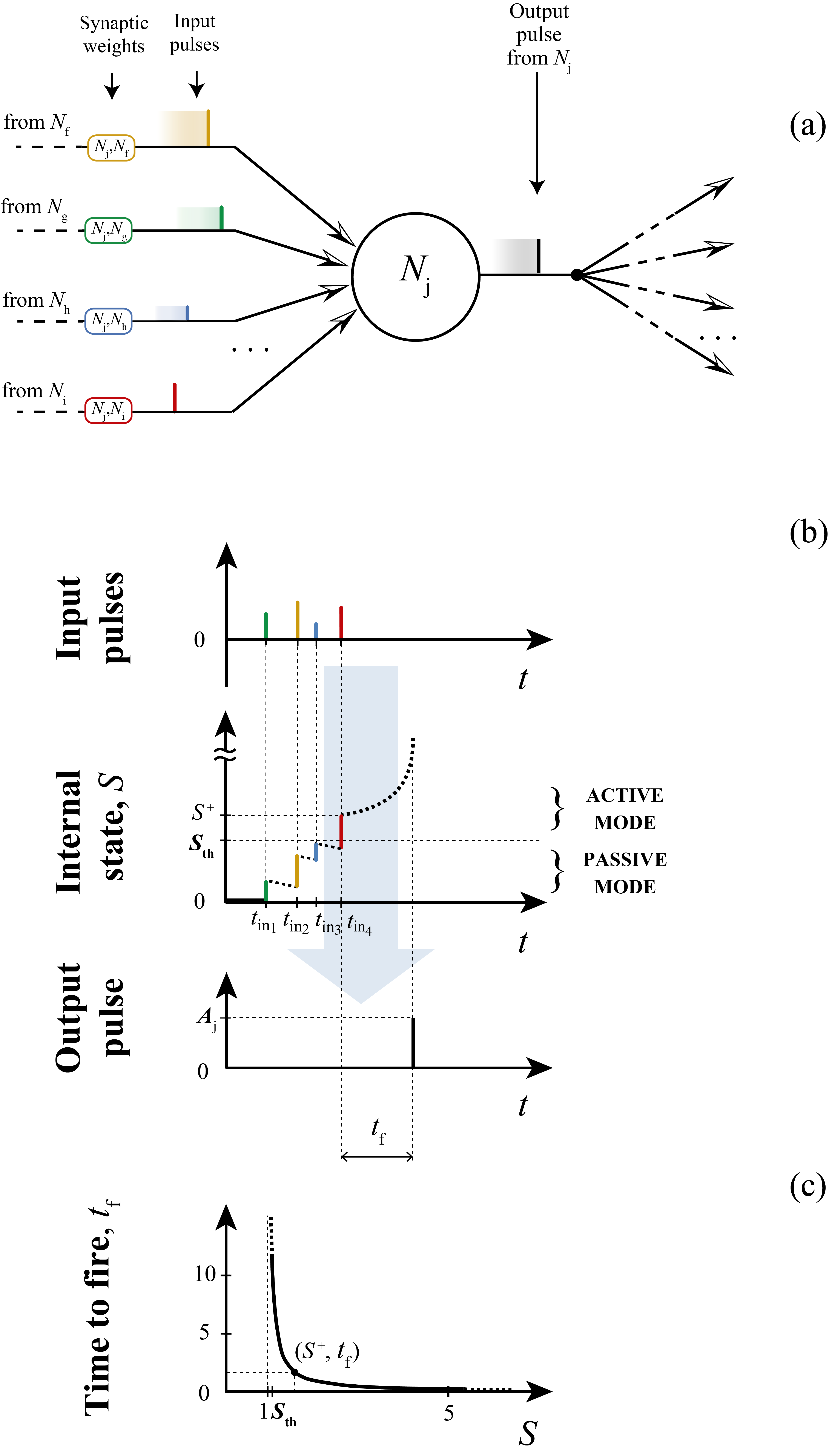}
 \caption{Neural summation and spike generation in a LIFL neuron. (a) Input/output process scheme; (b) temporal diagram of LIFL operation (basic configuration), assuming the neuron starts from its resting potential. For simplicity contributions are supposed to be all excitatory so that each incoming input causes an instantaneous increase of the internal state. In the passive mode the neuron is affected by a decay; when $S$ exceeds the threshold ($S=S^{+}$) the neuron is ready to spike; due to the latency effect, the firing is not instantaneous but it occurs after a time $t_{f}$. Once emitted, the pulse of amplitude $A_{Nj}$ is routed to all the subsequent connections, and then multiplied by the related weight. In (c) is shown the firing equation, i.e., the  latency curve for the determination of $t_{f}$ from $S^{+}$\citep[see][]{Salerno2011}. In this case $d$ is set to $0.04$}
\label{FIG:LIFLOp}
\end{figure}

$T_r$, the \emph{rise term}, takes into account the overthreshold growth of $S$ during two consecutive input pulses in the active mode. Specifically, once the neuron's internal state crosses the threshold, the neuron is ready to fire. However, firing is not instantaneous, but it occurs after a continuous-time delay. 
This delay time represents the spike latency, that we call \textit{time-to-fire}, and is indicated with $t_f$ in our model. 
$t_f$ can be affected by further inputs, making the neuron sensitive to changes in the network spiking activity for a certain period, until the actual spike occurs. 
$S$ and $t_f$ are related through the following relationship, called the \textit{firing equation}:
\begin{equation}
  t_f ={\frac{1}{(S-1)}} \; \label{EQ:Fe}
\end{equation}
Such dependence has been obtained through the simulation of a membrane patch stimulated by brief current pulses (0.01 \emph{ms} of duration) and solving the HH equations \citep{Hodgkin1952} in \emph{NEURON} environment \citep{Neuron}, as described in \cite{Salerno2011}.\\
The firing equation is a simple bijective relationship between $S$ and $t_{f}$, observed in most cortical neurons \citep{Izhikevich2004}; similar behaviors have been found by other authors, such as \cite{Wang2013} and \cite{Trotta2013}, using DC inputs. \\%Then, if the internal state of a neuron is known, the related $t_{f}$ can be calculated by means of Eq. \ref{EQ:Fe}.
The firing threshold is written as:
\begin{equation}
S_{th} = 1+d \; \label{EQ:Sth}
\end{equation}
where $d$ is a positive value called \emph{threshold constant}, that fixes a bound for the maximum value of $t_{f}$. 
According to Eq. \ref{EQ:Sth}, when $S = S_{th}$, $t_{f}$ is maximum, and equals to:
\begin{equation}\label{EQ:TfM}
t_{f,max} = 1/d
\end{equation}
$t_{f,max}$ represents the upper bound of the time-to-fire and is a meassure of the finite maximum spike latency of the biological counterpart \citep{FitzHugh1955}.\\
Under proper considerations (see sect.1 of \emph{Supplementary material}), it is possible to obtain $T_r$ (\emph{rise term}), as follows:
\begin{equation}
T_r =\frac{(S_{p}-1)^{2} \Delta t}{1-(S_{p}-1)\Delta t} \; . \label{EQ:Rt}
\end{equation}
$S_{p}$ represents the neuron's previous state, and $\Delta t$ is the temporal distance between two consecutive incoming presynaptic spikes. 
Eq. \ref{EQ:Rt} allows us to determine the internal state of the postsynaptic neuron at the time that it receives further inputs during the $t_f$ time window. 
In Fig. \ref{FIG:LIFLOp}, the operation of LIFL is illustrated. 
Neurons are supposed to interact instantaneously, through the \emph{synaptic weight} $w(N_j,N_i)$. 
Such link element can introduce amplification/attenuation of the traveling pulse.\\
The operation of the LIFL model is illustrated in Fig. \ref{FIG:LIFLOp}. 
Note that in this, and following figures, the synaptic weight is displayed with rounded rectangles, and identified by its post- and pre- synaptic neurons respectively.
For a given neuron $N_j$ operating in the active mode, the arrival of new input pulses updates the time-to fire $t_{f}$. If no other pulse arrives during this interval, the output spike is generated and $S$ is reset.\\
%Note that if incoming spikes are such as to bring $S$ to a value $<0$, $S$ is automatically set to 0 (a spike is immediately generated).
The presented basic configuration of the LIFL model defines an intrinsically \emph{class 1 excitable}, \emph{integrator} neuron, supporting \emph{tonic spiking} and \emph{spike latency}. 
We also included in the neuron model the \emph{absolute refractory period}, for which after the spike generation, the neuron's internal state remains at zero for a period $t_{arp}$, arbitrarily set. 
During this period the neuron becomes insensitive to further incoming spikes.

\subsection{STDP} 
\label{SECT:STDP}

STDP is a well-known type of plasticity mechanism consisting of an unsupervised spike-based process that can modify the weights on the basis of network activity. 
It underlies learning and information storage in the brain, and refines neuronal circuits during brain development \citep{Sjostrom2010}. 
The STDP mechanism influences the synaptic weights on the basis of the difference between the time at which the pulse arrives at the presynaptic terminal and the time a pulse is generated in the postsynaptic neuron.\\
\begin{figure}
\centering
\includegraphics[width=0.6\textwidth]{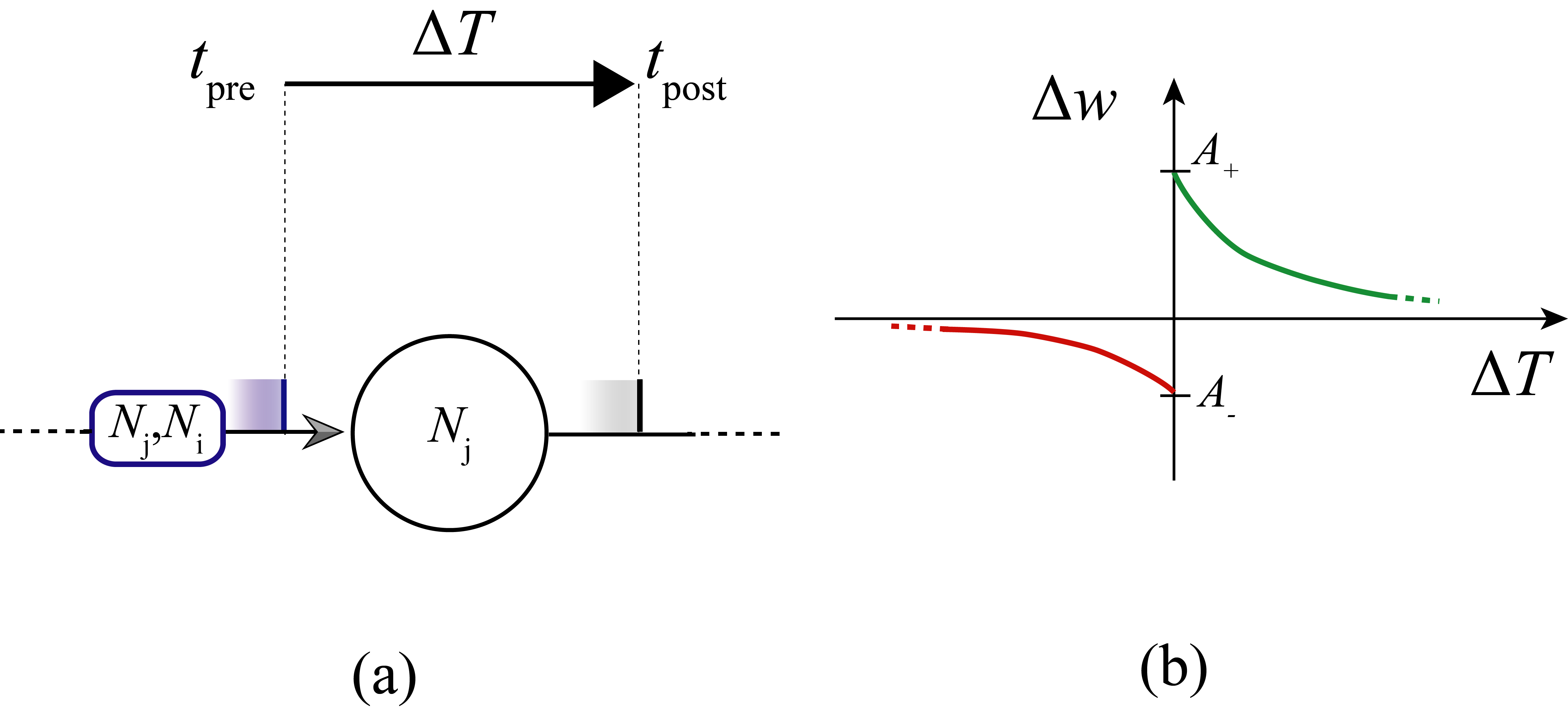}
\caption{Scheme of STDP. a) $\Delta T$; b) Learning window: LTD and LTP curves (in red and blue, respectively)}
\label{FIG:STDP}
\end{figure}
The original STDP behaviour \citep{Bi1998} can be modeled by two exponential functions \citep{Abbott2000}.
\begin{subequations}\label{EQ:STDP}
\begin{equation}
\Delta W =  A_+ \mathrm{e}^{-\frac{\Delta T}{\tau_+}} , \\ \mbox{ for } \Delta T > 0
\end{equation}
\begin{equation}
\Delta W = 0 , \mbox{ for } \Delta T = 0
\end{equation}
\begin{equation}
\Delta W =  A_- \mathrm{e}^{\frac{\Delta T}{\tau_-}} , \mbox{ for } \Delta T < 0
\end{equation}
\end{subequations}
where $\Delta T$ is the difference between the time at which the postsynaptic neuron fires (i.e., $t_{post}$) and the time at which the presynaptic pulse arrives (i.e., $t_{pre}$):
\begin{equation}
\label{EQ:Delta_t}
\Delta T = t_{post} - t_{pre}
\end{equation}
$\tau_+$ and $\tau_-$ are positive time constants for \emph{long-term potentiation} (LTP, Eq.\ref{EQ:STDP} a) and \emph{long-term depression} (LTD, Eq.\ref{EQ:STDP} c), respectively; $A_{+}$ and $A_{-}$ (positive and negative values, respectively) are the maximum amplitudes of potentiation and depression, that are chosen as absolute changes, as in other works \citep[e.g.,][]{Acciarito2017}. 
Then, a weight is increased or decreased depending on the pulse order (\emph{pre-}before \emph{post-}, or \emph{post-} before \emph{pre-}, respectively).\\
In this work we will focus on heterosynaptic STDP plasticity, by which the time difference between output and input pulses determines the modification of other synaptic afferents to the neuron.\\
%Heterosynaptic STDP [Hiratani et al. - Detailed Dendritic Excitatory/Inhibitory] have been studied for... \\ Relative differences in spike timings experienced among neighboring glutamatergic and GABAergic synapses on a dendritic branch significantly influences changes in the efficiency of these synapses; This heterosynaptic form of spike-timing-dependent plasticity (STDP) is potentially important for shaping the synaptic organization and computation of neurons,; GABA-driven neural circuit formation. [Hiratani et al. - Detailed Dendritic Excitatory/Inhibitory]. [Note that in the immature brain, GABA is excitatory and then becomes inhibitory..]''

\subsection{Multineuronal spike sequence detector}
\label{SECT:MNSD}

A broad range of literature is aimed at understanding how animals have the capability to learn external stimuli and to refine its internal representation. 
Many of these studies propose architectures based on delays and coincidence detection mechanisms \citep{Hedwig2017, Konig1996}\\
In a classic pattern recognition problem an object can be described by a n-dimensional vector (or matrix) where each component represents an object's feature. 
Analogously, in the neural computation context, an object can be identified by an n-uple of pulses, where feature attributes are encoded in the times at which the pulses occur \citep{Susi2015}. 
This allows us to map the classes in a n-dimensional topological space of the internal object representation.\\
We present here a multineuronal spike pattern detector that includes a bio-plausible self-tuning mechanism, that is able to learn and recognize multineuronal spike sequences through repeated stimulation, without supervision. 
We term this neuromorphic structure as \emph{Multi Neuronal spike-Sequence Detector} (MNSD). 
Through a MNSD we can mediate the mapping from spatio-temporal stimuli to such temporal feature space, identifying a class with a specific area, that we call \emph{class hypervolume}. 
In this section we show the operation principles on which such structure is based.

\begin{figure}
\centering
\includegraphics[width=0.5\textwidth]{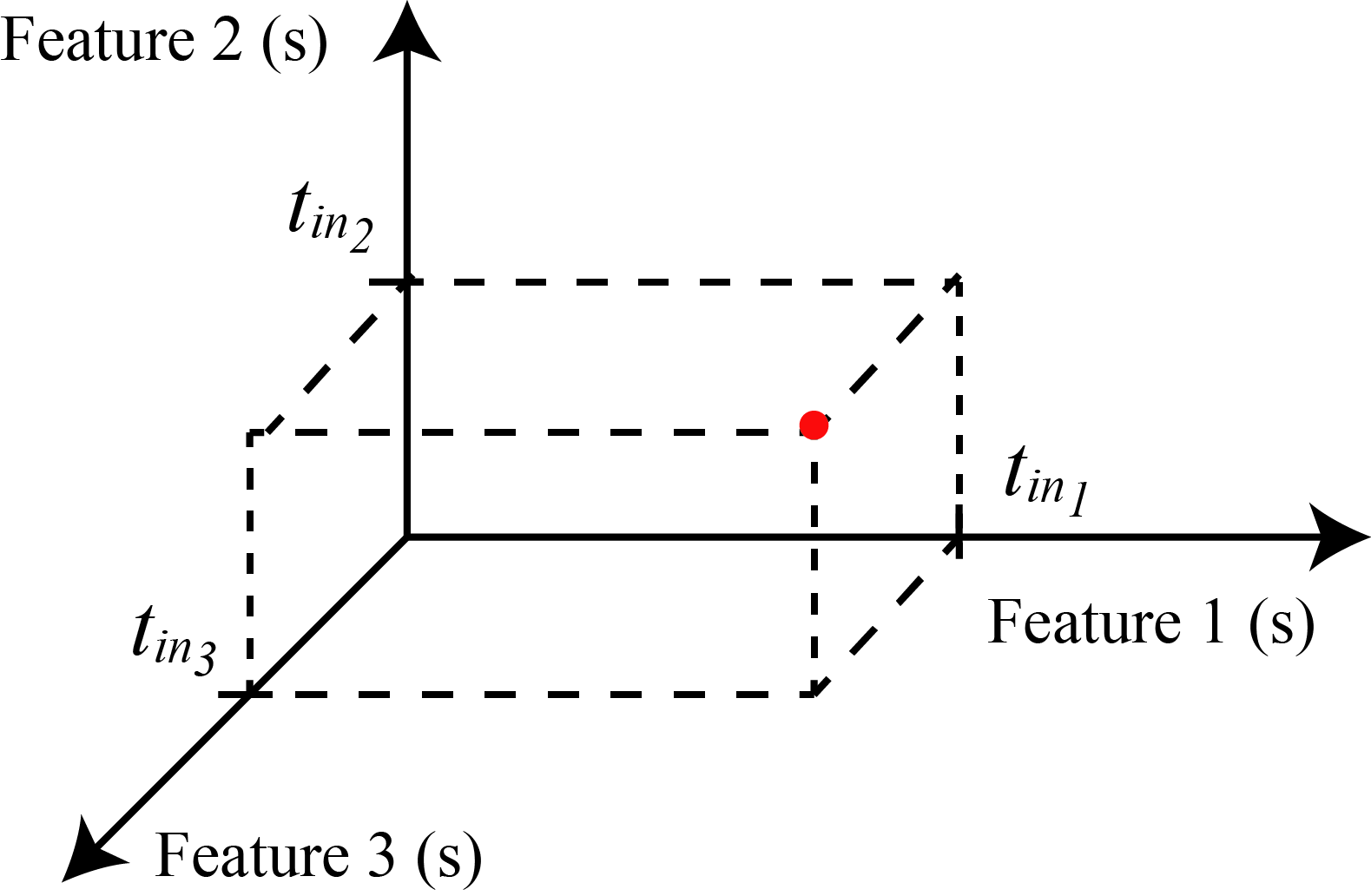}
\caption{An object characterized by three features can be identified in a three-dimensional feature space by the arrival times of three input pulses. In this way, given a multi neuronal spike sequence as input, the MNSD will associate it to the represented class whenever the input spikes fall in proper temporal ranges.}
\label{FIG:Space}
\end{figure}

\subsubsection{Structure description}
\label{SECT:StrDes}
\begin{figure}
\centering
\includegraphics[width=0.5\textwidth]{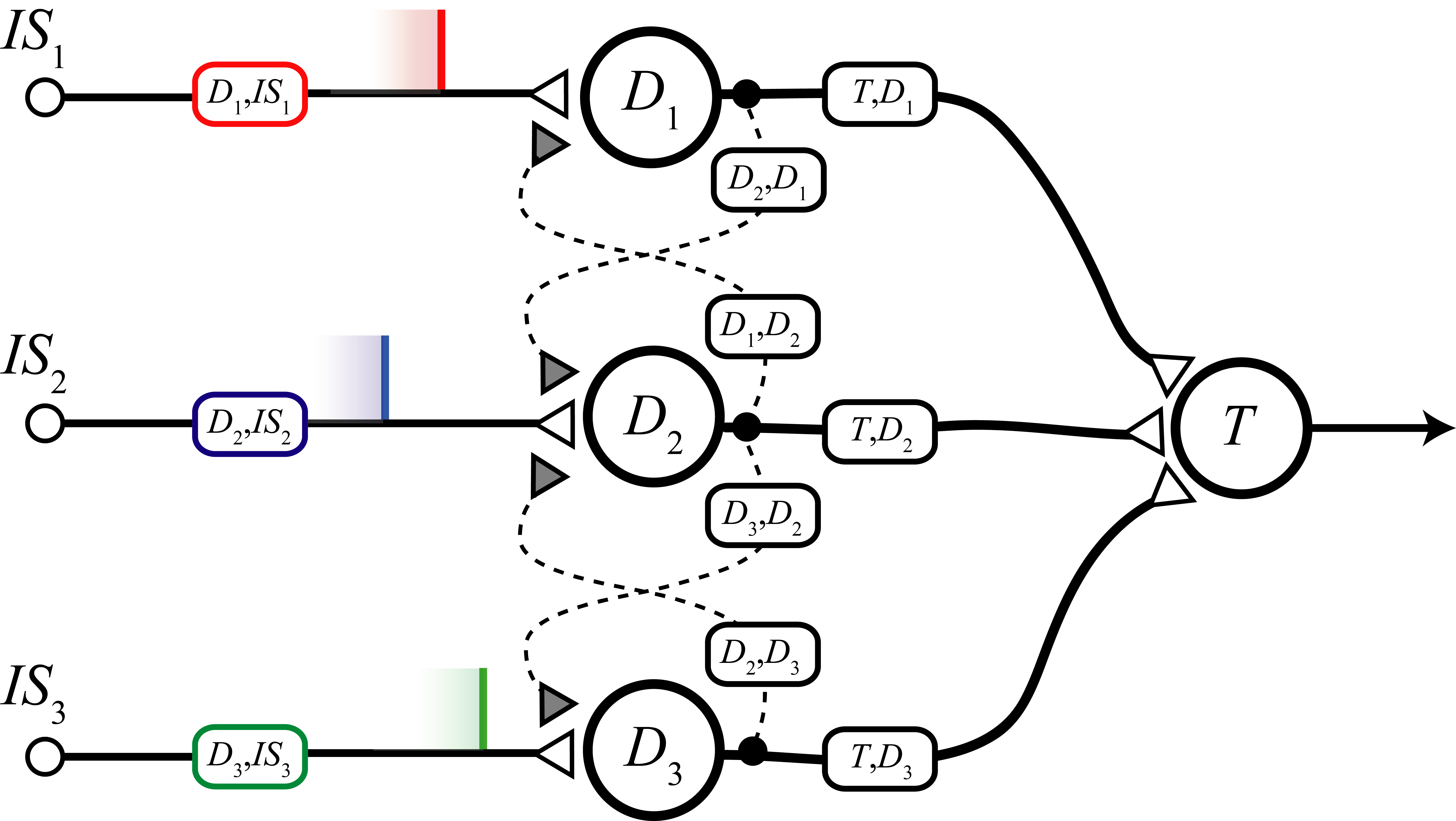}
\caption{Scheme of the presented structure. The three delay branches, characterized by the three delay neurons $D_1$, $D_2$, $D_3$, converge to the target neuron $T$. Heterosynaptic STDP interactions are permitted by lateral connections, represented as dotted curves with related synapses}
\label{FIG:structure}
\end{figure}

The MNSD architecture, represented in Fig. \ref{FIG:structure}, is composed of:

\begin{itemize}
\item a layer of neurons $D_1, ... , D_n$ (termed \emph{delay neurons}) that receive inputs spikes, 
%(each of which represents a single spike or a spike cluster)
and are subject to heterosynaptic STDP interactions between them. 
For simplicity we only consider nearest-neighbor interactions between delay neurons, i.e., each branch can interact with its neighbors branches only (in order to mimic a layer of adjacent neurons).
\item one target neuron $T$, that performs the summation of previous contributions and acts as readout neuron, signaling the recognition of the sequence. 
\end{itemize}

To facilitate the analysis and to map the spatio-temporal stimuli in three dimensions we will consider a structure with only three branches; nevertheless, we can generalize to structures of as many branches as features of the object we want to classify. 
We also consider that the interactions between the neurons are instantaneous; then the only possible delays in the network are those produced by the spike latency. 

In order to perform the recognition, the structure's weights $w(D_n, IS_n)$ are adaptively adjusted on the basis of the specific mutineuronal spike sequence given at the input. In this way the target neuron ($T$) will become active only at the presentation of such sequence (or similar ones, as we will see in sect. \ref{SECT:DynamicalAnalysis}).\\
The necessary condition for $T$ to spike is that $S>S_{th}$; this is made possible by the synchronization of the (excitatory) contributions coming from the delay branches. Synchronizability at the target neuron in response to the specific sequence is progressively obtained through the repeated presentation of the sequence to the structure, due to the interplay between the spike latency and the heterosynaptic STDP. Through the amplitude-time transformation operated by the spike latency feature it is possibile to obtain synchronization on the target neuron acting on the amplitude of the pulses at the input of the delay neurons. The spike latency feature is then fundamental for the correct operation of the structure (a simple LIF would not be able to support this mechanism). The interaction between adjacent branches (lateral excitation) combined with the hetherosynaptic STDP make it possible $w(D_n,IS_n)$ to change with respect to the difference between their spike times. This modifies the amplitudes of the contributions in the input of the different branches, enabling a feedback mechanism to mutually compensate the differences between the output spike times of adjacent branches and to produce a synchronous arrival to the target. 

%In this way, structures are permitted to auto-tune on particular ISIs, in order to detect the presented sequences.

With the aim of better explaining the operation of the MNSD structure, we initially perform an analysis of the structure without plasticity (i.e., \emph{static analysis}). 
Later, we will include a (hetero-)synaptic term to show how one branch can adapt dynamically to reach a downstream spike synchrony with its neighbor (\emph{dynamical analysis}). 
In order to design structures that are capable to face real problems by operating with this principle, we will derive the set of relations in sects.\ref{SECT:StaticAnalysis} and \ref{SECT:DynamicalAnalysis}, and then we tune a MNSD for a specific application (sect. \ref{SECT:Results}).

\subsubsection{Static analysis}
\label{SECT:StaticAnalysis}
\begin{figure}
\centering
\includegraphics[width=0.8\textwidth]{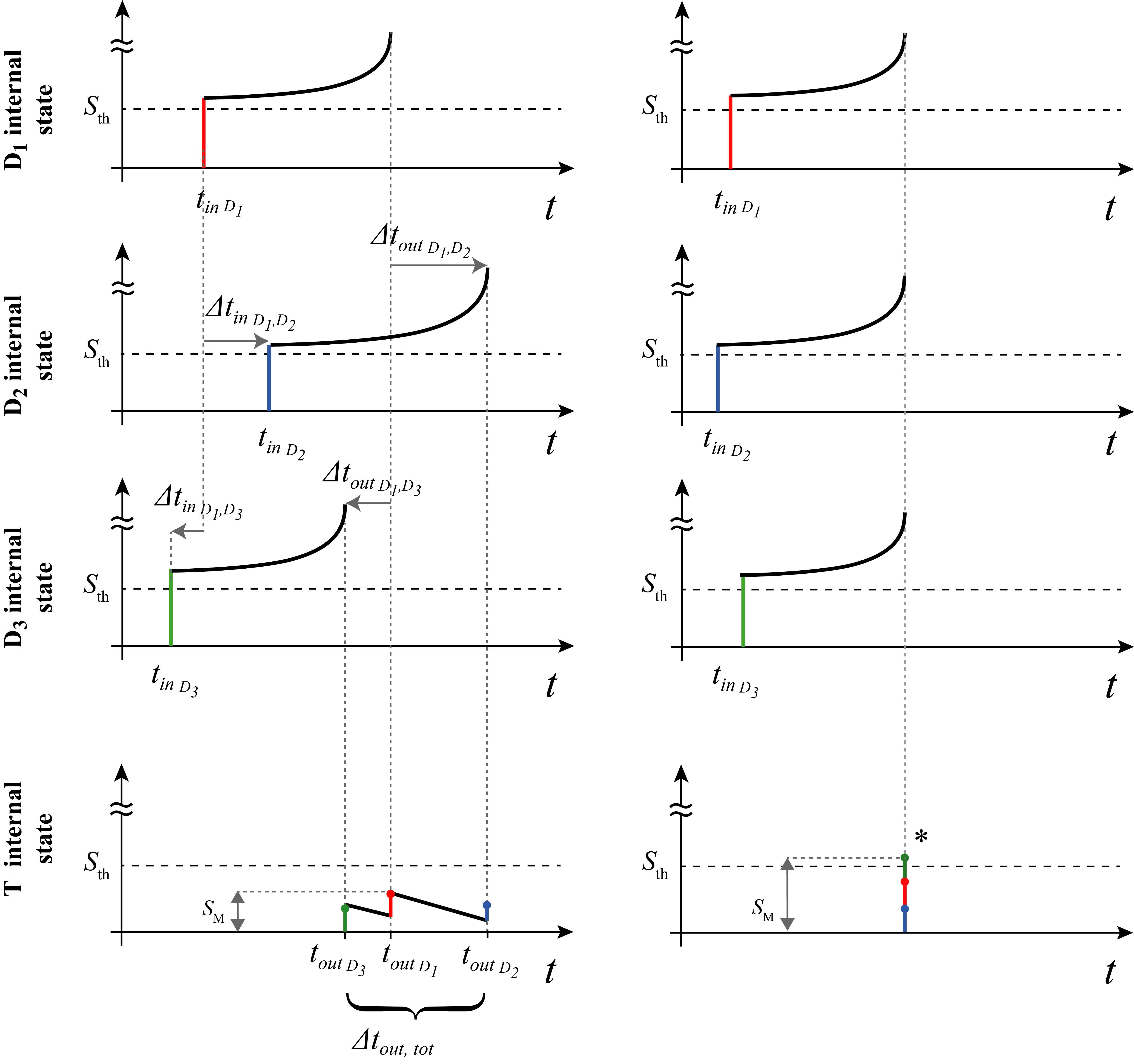}
 \caption{Diagram of the operating principle of the structure (static analysis). On the left, desynchronized input pulses are unable to activate the target. Note that depending on the arrival orders of $t_{in}$ (and $t_{out}$), some $\Delta_{in}$ (and $\Delta_{out}$) can assume negative values (arrow directions are significant). At right, \emph{simultaneity condition} allow the target activation. Finally, maximum state $S_M$ is represented for both the cases}
\label{FIG:diagram}
\end{figure}

In this section we obtain the conditions that allow $T$ to generate a spike, without considering the plastic term (i.e., not considering the dotted connections of Fig.\ref{FIG:structure}). 
The operation of the structure in the static mode is shown by means of the temporal diagrams in Fig.\ref{FIG:diagram}. \\ % Per iniziare la nostra analisi, supponiamo che i rami siano in condizioni tutte uguali. 
The excitatory neurons $D_n$ , present in the afferent branches, allow to create a transmission delay through the spike latency mechanism. The operation is based on the fact that the pulses arriving from the different branches can evoke a spike in $T$ only if they  arrive sufficiently close in time.\\
In the following we indicate with $t_{in_{D_n}}$ the arrival instant of the input spike $IS_n$ and with $t_{out_{D_n}}$ the time at which the output pulse of $D_n$ is generated; $\Delta t_{in_{D_m,D_n}}$ represents the time difference between the pulses afferent to the delay neurons (i.e., $t_{in_{D_n}}-t_{in_{D_m}}$), and $\Delta t_{out_{D_m,D_n}}$ the time difference between the pulses afferent to the target neuron (i.e., $t_{out_{D_n}}-t_{out_{D_m}}$).
Let us consider the amplitude of the pulses. At the input, and to guarantee the activation of $D_{n}$, the following relation has to be satisfied:
\begin{equation}
\label{EQ:cond1}
A(IS_{n})\cdot w(D_n, IS_n)\ge 1+d
\end{equation}
where $A(IS_{n})$ is the amplitude of the input pulse, $w(D_{n}, IS_n)$ the synaptic weight afferent to $D_{n}$, and their product represents the amplitude of the input pulse arriving to $D_n$. For simplicity we consider that:
\begin{itemize}
\item neurons are identical, i.e., initialized with the same set of parameters presented in sect.\ref{SECT:NeuMod};
\item synaptic weights afferent to the target are the same for the three afferent connections: \\ 
$w(T,D_1)=w(T,D_2)=w(T,D_3)=w(T,D_n)$ 
\item External spikes $IS$ , as well as output pulses, are assumed to have the same amplitude ($A(IS_{n})=1$)
\end{itemize}
Then:
\begin{equation}
\label{EQ:cond3}
w(D_n) \ge 1+d
\end{equation}
%\begin{equation}
%\label{EQ:cond2}
%\frac {1}{w(D_n)-1}\ge \frac {1}{d}
%\end{equation}
%dove abbiamo usato la \ref{EQ:Fe} ambo i membri. Then:
%                       [SETTING PESI AL TARGET]
Assuming that the pulses arrive simultaneously at the target (simultaneity condition), we have that the following relation has to be satisfied to guarantee the output spike of neuron $T$:
\begin{equation}
\label{EQ:cond4}
w(T,D_{n}) \ge \frac {1+d}{3}
\end{equation}
In order to have the target activated with the contribution of all the three branches (avoiding that the target neuron generates a spike also for partial sequences that do not exhibit the whole set of features of our object), we have the following constraint:
\begin{equation}
\label{EQ:cond6}
w(T,D_n)<\frac{1+d}{2}
\end{equation}

%Then it is possible to merge the previous conditions, as follows:

%\begin{equation}
%\label{EQ:cond3}
%DA_SISTEMARE % P(TN,DN) ]- Ld|1/((〖Iext〗_2-1) )–1/(〖Iext〗_1–1)  –1/[P_r⁡〖P_w (DN,〖EI〗_1 )〗-1] +〖∆t〗_in |>(1+d)
%\end{equation}

Now we introduce the delay times due to the spike latency. Considering Fig.\ref{FIG:diagram}, we can write the system of equations that relates the arrival times of the three contributions to $T$ as:

\begin{subequations}
\label{EQ:sist1}
\begin{equation}
t_f (D_1) + \Delta t_{out_{D_1,D_2}} = \Delta t_{in_{D_1,D_2}} + t_f (D_2) 
\end{equation}
\begin{equation}
t_f (D_1) + \Delta t_{out_{D_1,D_3}} = \Delta t_{in_{D_1,D_3}} + t_f (D_3)
\end{equation}
\end{subequations}
%OK

In order to achieve simultaneous arrival of the pulses to the target, we should have:

\begin{equation}
\Delta t_{out_{D_m,D_n}} = 0.  
\end{equation}

Then:

\begin{subequations}
\label{EQ:sist2}
\begin{equation}
t_f (D_1) =  \Delta t_{in_{D_1,D_2}} + t_f (D_2)
\end{equation}
\begin{equation}
t_f (D_2) = \Delta t_{in_{D_1,D_3}} + t_f (D_3)
\end{equation}
\end{subequations}

This means that, for a  simultaneous arrival of pulses at the target, with the above-mentioned restrictions, we should have:

\begin{subequations}
\label{EQ:sist3}
\begin{equation}
\Delta t_{in_{D_1,D_2}} = \frac{1}{w(D_1, IS_1)-1} - \frac{1}{w(D_2, IS_2)-1}
\end{equation}
\begin{equation}
\Delta t_{in_{D_1,D_3}} = \frac{1}{w(D_1, IS_1)-1} - \frac{1}{w(D_3, IS_3)-1}
\end{equation}
\end{subequations}

Now we remove the simultaneity condition at the target, searching for the values of $\Delta t_{in}$ and $w(T, D_n)$ for which the spike at the target neuron is still allowed. Under proper considerations (see sect.2 of \emph{Supplementary material}), we arrive to the following relations:\\
If $\Delta t_{in_{D_1,D_2}}$ and $\Delta t_{in_{D_1,D_3}}$ have concordant sign, then:
\begin{align}
& max (|\Delta t_{in_{D_1,D_2}}-\frac{1}{w(D_1, IS_1)-1}+\frac{1}{w(D_2,IS_2)-1}|,\nonumber \\ 
& \quad |\Delta t_{in_{D_1,D_3}}-\frac{1}{w(D_1, IS_1)-1}+\frac{1}{w(D_3, IS_3)-1}|) <\frac{2-d}{L_d}
\end{align}
On the contrary, if $\Delta t_{in_{D_1,D_2}}$ and $\Delta t_{in_{D_1,D_3}}$ have discordant sign, then:

\begin{align}
&|(\Delta t_{in_{D_1,D_2}}-\frac{1}{w(D_1, IS_1)-1}+\frac{1}{w(D_2, IS_2)-1}) - \nonumber \\
& \quad (\Delta t_{in_{D_1,D_3}}-\frac{1}{w(D_1, IS_1)-1}+\frac{1}{w(D_3, IS_3)-1})| <\frac{2-d}{L_d}
\end{align}

If we aim at recognizing parallel spike trains of greater cardinality, it is necessary to increase the number of delay branches, keeping the condition that the contributions have to arrive simultaneously to the target neuron.\\

\subsubsection{Dynamical analysis}
\label{SECT:DynamicalAnalysis}

\begin{figure}
\centering
\includegraphics[width=0.4\textwidth]{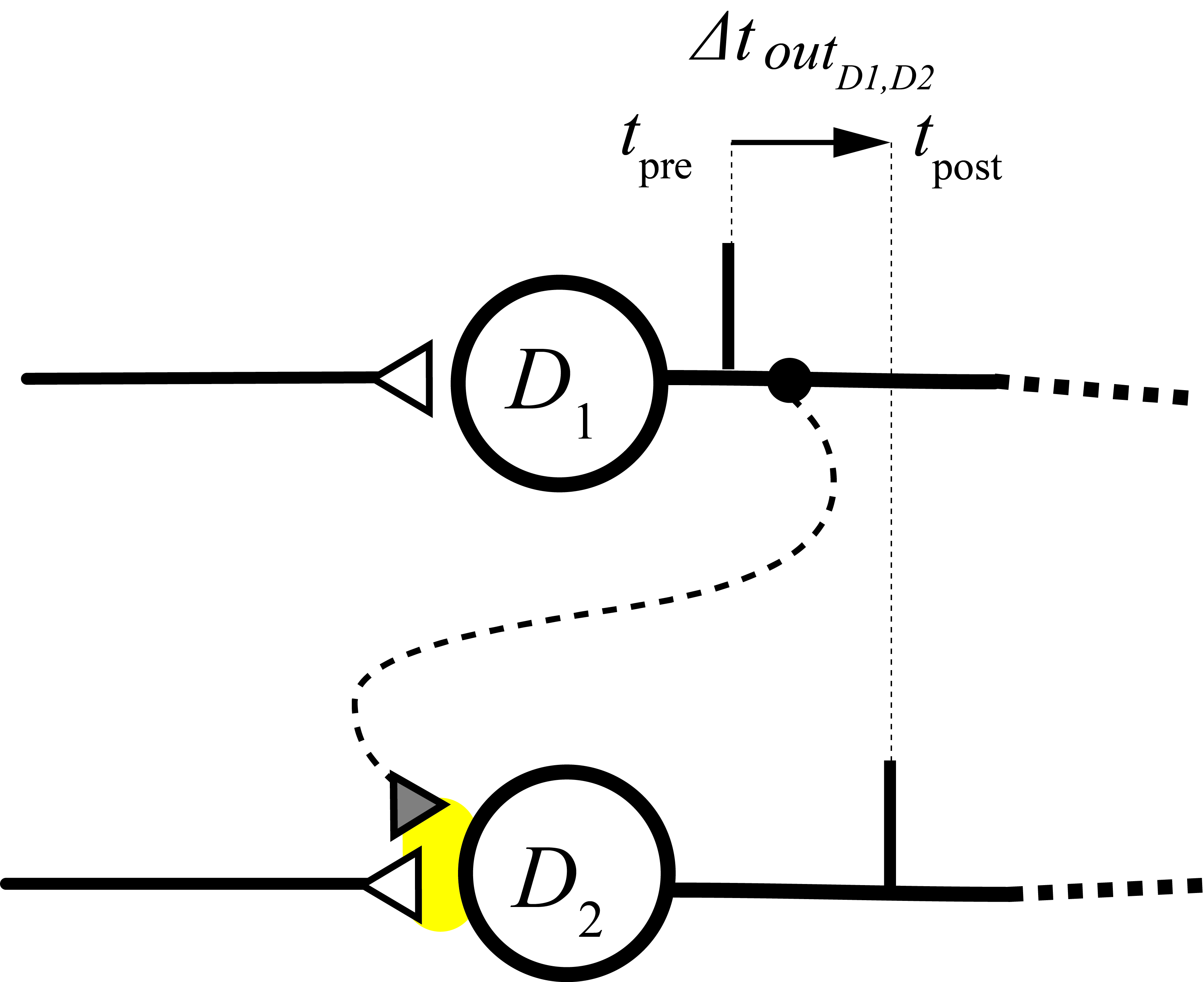}
 \caption{Lateral excitatory heterosynaptic junction. The area of synapse modification is highlighted in yellow}
\label{FIG:heterocouple}
\end{figure}

As already mentioned, the operational key of the structure resides in the interplay of spike latency and plasticity: the delay in neuronal pathways is due to the spike latency, which in turn depends on $w(D_n, IS_n)$. 
In addition $w(D_n, IS_n)$ is modulated by the neighbor branch(es) through heterosynaptic plasticity. 
Therefore, the branch delay is modulated by plasticity.
In the presence of plasticity and under repetitive stimulation, the structure can progressively self-regulate its weights until the multineuronal spike train synchronizes in the target neuron (operation mode described in the previous section).\\
For simplicity, and without loss of generality, we consider here the effect of a single heterosynaptic connection (the influence of a single branch on an adjacent one). 
In the whole structure, however, each branch acts on its neighbors through heterosynaptic lateral junctions. 
This leads to a modification of the timing of the branch's pulse in order to converge to the neuron $T$ temporally closer with respect to their neighbor(s). 
Such mechanism is shown in Fig.\ref{FIG:heterocouple} where heterosynapsis is indicated with a dotted curve and a grey triangle. 
In this way the weight $w(D_2, IS_2)$ is modulated by the time difference between the output pulse of $D_2$ and the contribution from $D_1$ (i.e., the output pulse of $D_1$). 
In the case of generic heterosynaptic plasticity, the weight potentiation/depression will also involve the other afferences of $D_2$, but if we assume the lateral contribution to be weak, both its contribution (see Eq.\ref{EQ:Rt}) and the weight variation will be fall, while the modifications related to the input $w(D_2,IS_2)$ will be affected in a sensible manner. 
For simplicity, we will consider heterosynaptic modulation, acting on $w(D_2,IS_2)$ only.

Considering that connections are instantaneous (as specified in sect. \ref{SECT:MNSD}), we note that the $\Delta T$ cited in sect. \ref{SECT:STDP} in this configuration corresponds to $\Delta t_{out{D_1,D_2}}$ (see Fig.\ref{FIG:heterocouple}). 
We can then write:

\begin{subequations}
\begin{equation}
\Delta w(D_2, IS_2) =  A_+ \mathrm{e}^{-\frac{\Delta t\:out_{D_1,D_2}}{\tau_+}} , \mbox{ for } \Delta t\:out_{D_1,D_2} > 0
\end{equation}
\begin{equation}
\Delta w(D_2, IS_2)  = 0 , \quad \quad \quad \quad \quad \quad \mbox { for } \Delta t\:out_{D_1,D_2} = 0
\end{equation}
\begin{equation}
\Delta w(D_2, IS_2) =  A_- \mathrm{e}^{\frac{\Delta t\:out_{D_1,D_2}}{\tau_-}} , \mbox{ for } \Delta t\:out_{D_1,D_2} < 0
\end{equation}
\end{subequations}

\begin{figure}
\centering
\includegraphics[width=0.7\textwidth]{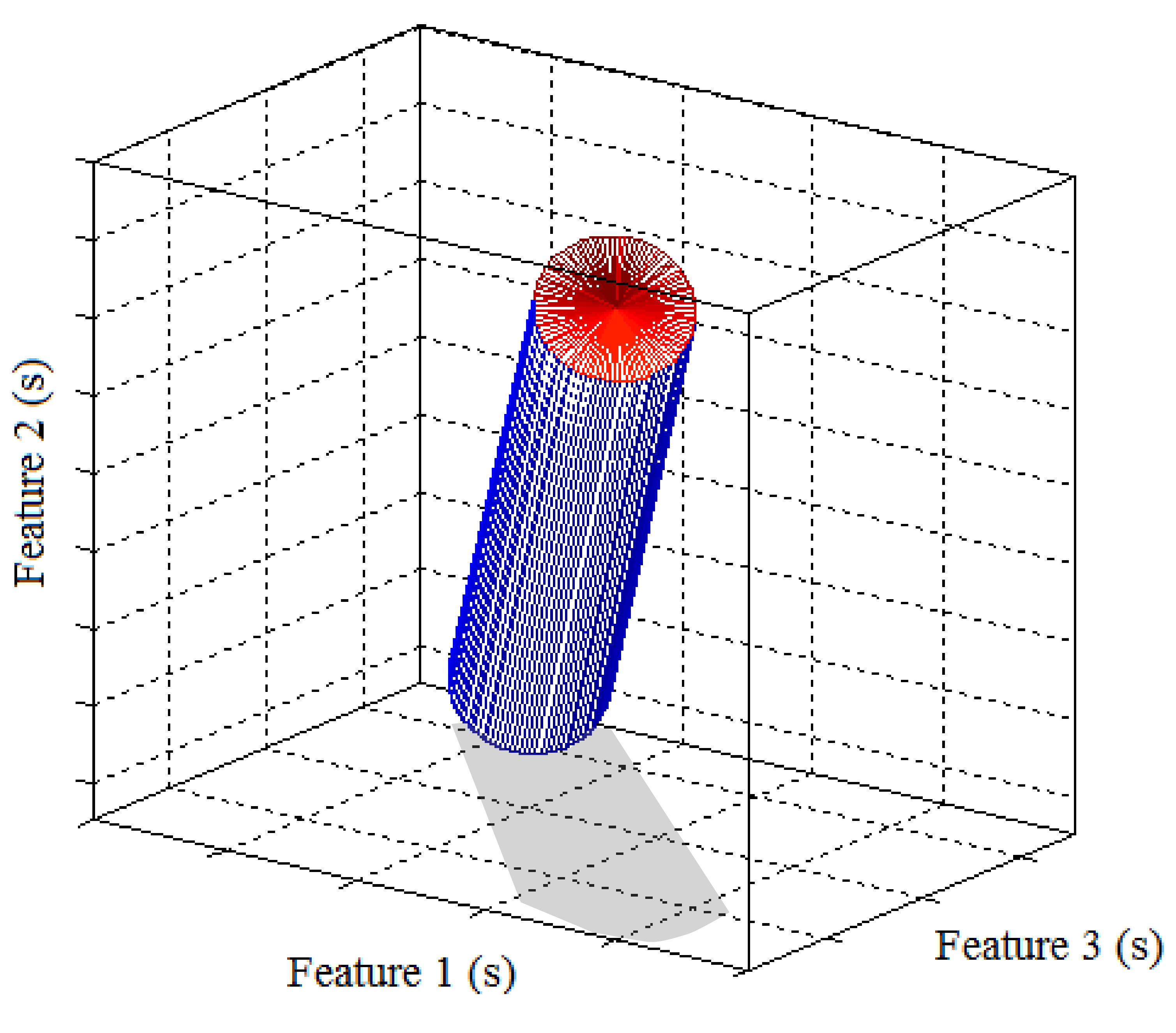}
 \caption{Representation of class hypervolume identified by a three dimensional MNSD. While STDP translates the axis of the cylinder,  $L_d$ varies its radius. For this figure we assumed that the multineuronal patterns arrive to the MNSD when the neurons are at their resting potentials}
\label{FIG:3DClassHypervolume}
\end{figure}

The difference $\Delta t\:out_{D_1,D_2}$ elicits an increase of the weight $w(D_2, IS_2)$ when the arrival pulses order is $D_2$, $D_1$ (a decrease otherwise), causing a decrease (increase) of the latency at the arrival of the next $IS_2$.
Synaptic changes must be induced by spikes belonging to the same sequence.
Consequently, it is important to prevent interference between subsequent multineuronal sequences.
This is done by carefully adjusting the STDP time constants.\\
In some scenarios, we aim at a certain tolerance to a temporal jitter of the input spikes. 
By changing the decay constant $L_d$ we can modulate the tolerance of the structure leading to a stronger selectivity (robustness) to the jitter present in input patterns.
The higher (lower) the $L_d$, the more selective (robust) the structure becomes to the jitter.
%, corresponding to a smaller (bigger) \emph{class hypervolume}), allowing the target to turn on only for new sequences which spikes sequences are similar to those used for the training\\
Another relevant characteristic is that, when using the MNSD, the detection does not depend on the arrival time of the first spike but only on the intervals between spikes. 
%This allows identifying patterns in a continuous data stream.
In a three dimensional feature problem (characterized by three neuronal pathways), the corresponding hypervolume is (in our case where all $L_d$ are equal) a cylinder whose radius depends on $L_d$ and its axis $\zeta$ has a slope of 45$^{\circ}$ with respect of each of the coordinates (see Fig. \ref{FIG:3DClassHypervolume}).
Its mathematical form is defined by the following expression:

\begin{equation}
\label{EQ:ClassHypAxis}
\zeta = (t_{offset}+\frac{1}{w(D_1, IS_1)-1}, t_{offset}+\frac{1}{w(D_2, IS_2)-1}, t_{offset}+\frac{1}{w(D_3, IS_3)-1})
\end{equation}

Where $t_{offset}$ is the time of arrival of the first pulse of the sequence.
In Fig. \ref{FIG:3DClassHypervolume} we  represent the cylinder defined by our MNSD. 
If the arrival times of a pattern fall into the cylinder, the MNSD produces a spike.

\section{Results}
\label{SECT:Results}
\begin{figure}
\centering
\includegraphics[width=0.5\textwidth]{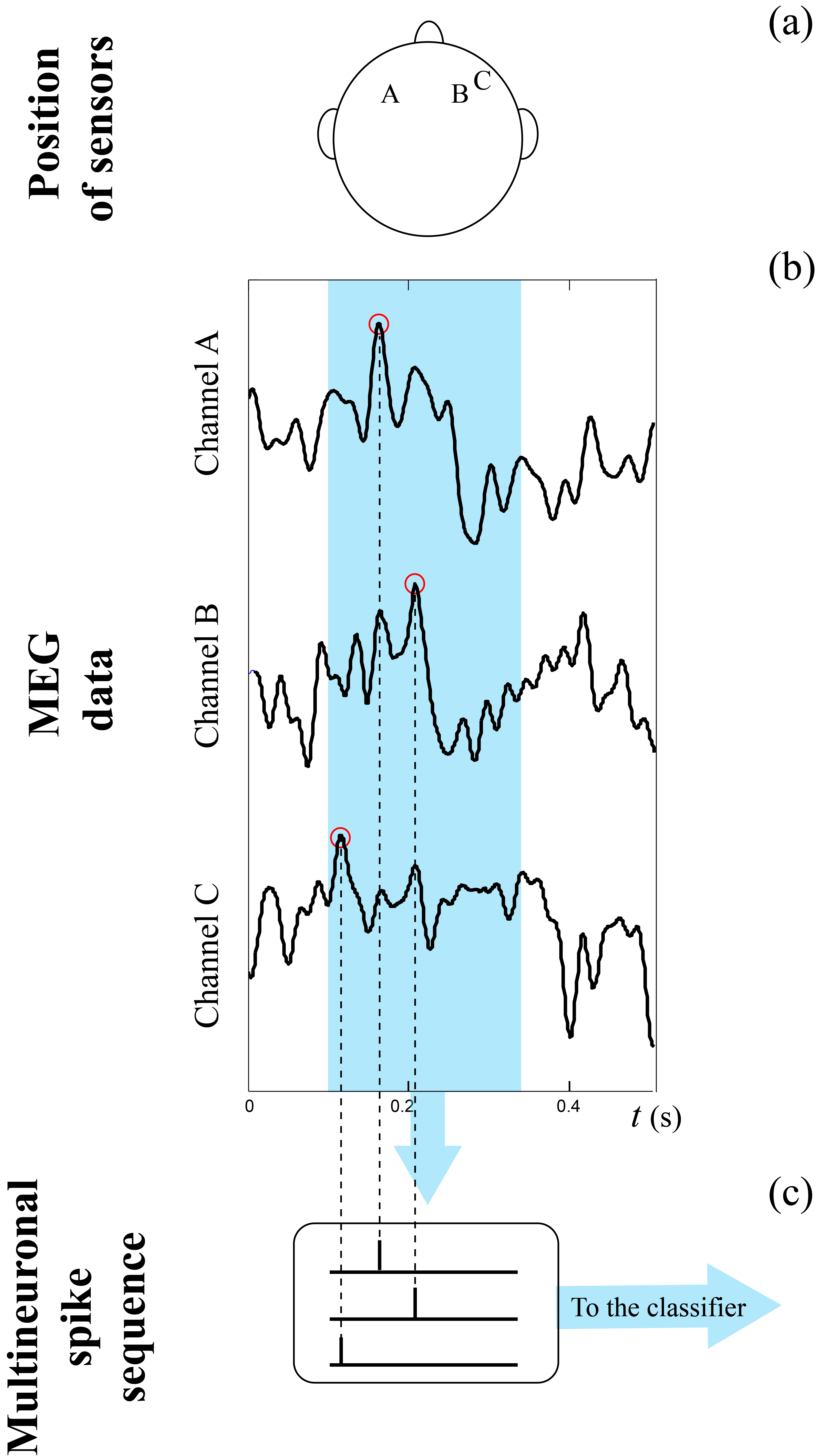}
\caption{Multineuronal spike sequence generation process. (a) Position of sensors: channel A (left prefrontal region), channel B and channel C (right prefrontal region); (b) extraction of the time series from a single trial: only signals deriving from three channels are considered; (c) maxima are selected to generate multineuronal spike sequences}
\label{FIG:spiketrains}
\end{figure}

In order to show how the developed MNSD tool can be used to study pattern recognition problems, we implemented the structure to perform the recognition of cognitive states, using real data from a motor-inhibitory (Go/NoGo) task \citep{Falkenstein1999, LopezCaneda2017}. Such paradigm is useful to study neural substrates of response inhibition and sustained attention processes. 
\emph{Event-related potentials} studies have found discriminative neuroelectric components (e.g. $N2$ and $P3$, \citep{Eimer1993, Falkenstein1999, Falkenstein2006}) between target and non-target conditions, evidencing inhibition functional networks and different motor responses \citep{Lavric2004, Kamarajan2004, Pandey2012}.\\
The two classes of stimuli have been presented to 67 participants (age range: 13-15 years old), were blue squares/green circles as targets (Go) and green squares/blue circles as non-targets (NoGo), displayed randomly and with an equiprobable presentation ratio. 
Participants were instructed to press a button as fast as possible only when a target was shown in the center of the screen (with the right hand Go and the left hand for NoGo). 
The stimuli were presented for 100 ms with a \emph{stimulus onset asynchrony} (time interval between two trials) of $1400 \pm 200$ ms.\\
High-density MEG signals were obtained from 306 channels (102 pairs of planar gradiometers and 102 magnetometers) with an Elekta Neuromag Vectorview system situated in a magnetically and electrically shielded room. 
Only the 102 Magnetometers were used to carry out the analysis. 
The signals were recorded with a 1000 Hz sampling rate and filtered online with a band pass 0.1-330 Hz filter. 
A \emph{3Space Isotrak II} system was used for the registration of the magnetic coil positions, fiduciary points and several random points spread across the participant scalp. 
For this preliminary study, we have chosen randomly one of the participants that performed this task 
% (a 14 years old healthy subject)
and considered a total of 150 trials for the dataset.\\
Methods were carried out in accordance with the approved guidelines and general research practice. The study was approved by the ethical committee of the Complutense University of Madrid. Informed consent has been obtained from the parents (or guardians) of the subjects, since they are under the age of 16.\\
Although a statistical test revealed clear differences between the two conditions on a sufficiently large set of samples, neural noise makes the trial-specific discrimination between the two classes of responses not trivial. 
To overcome this limitation, we extracted in each trial the segment in the time interval $[0.1, 0.35] s$ after the stimulus presentation, to avoid the premotor response (which starts around 400 ms) \citep{Deecke1976, Ikeda2000}. 
This reduces artifacts and ensures that the activity is related to the cognitive task and not to the motor action.  
Then, we performed a second statistical test to select those channels whose time series exhibit clear differences between the two response classes. 
In this way we selected the three representative sensors $0341$, $1221$ and $1411$ (that we call channel A, channel B and channel C, respectively) as the most significant ones.
Such sensors are located in prefrontal regions (see Fig. \ref{FIG:spiketrains}a), which agrees with the literature of the field since prefrontal regions have been associated with inhibitory cognitive responses \citep{Chambers2009}.
From the time series of these channels we extracted the maximum peaks (Fig. \ref{FIG:spiketrains}b) and transformed them into spike sequences (Fig. \ref{FIG:spiketrains}c).\\
We realized a classifier based on a single MNSD trained to recognize the distinctive timings of the Go class, considering 70\% of the used dataset for the learning (105 Go samples). 
For the test phase we used both Go and NoGo samples (23 Go and 22 NoGo). 
During the training phase the structure adjusted its weights due to plasticity effects while during the test phase the weights were kept constant. 
The target neuron produced a spike only when the Go class was detected, allowing us to differentiate between Go and NoGo classes.
In order to set up the MNSD, we implemented equations in \ref{SECT:StaticAnalysis} and \ref{SECT:DynamicalAnalysis} in Matlab\textcircled R environment. 
Taking into account the constrains for the correct operation of the MNSD, and to make the structure compatible with this problem, we initialized it as follows:
\begin{itemize}
\item $t_{f,max}$ larger than the maximum possible $\Delta t_{in}$. To achieve it, we reduced the $250$ ms interval of the segments by a factor 10, obtaining sequences of 25 ms, and set $c=0.04$ (i.e., $t_{f,max}$ = $25ms$);
\item We chose input amplitudes that led $D_n$ around the center of the latency range (i.e., $t_{f,max}/2$ =$12.5ms$), to obtain the largest margin to set $D_n$. To achieve it, we set $A_n=1$ and $w(D_n, IS_n)=1.08$;
\item $L_d$ was chosen sufficiently low (equals to $0.15$) to have a tolerant structure, since we are dealing with a noisy scenario;
\item For the STDP we limited $\tau_+$ and $\tau_-$ in a range that avoids interaction between adjacent sequences; it is also useful to take $A_+$ and $A_-$ in a range where abrupt changes of weight values are avoided while the presentation of the patterns. We set $A_+= 0.0035$; $A_-= -0.0035$; $\tau+= 10$; $\tau_-=10$. 
\end{itemize}

\begin{figure}
\centering
\includegraphics[width=0.7\textwidth]{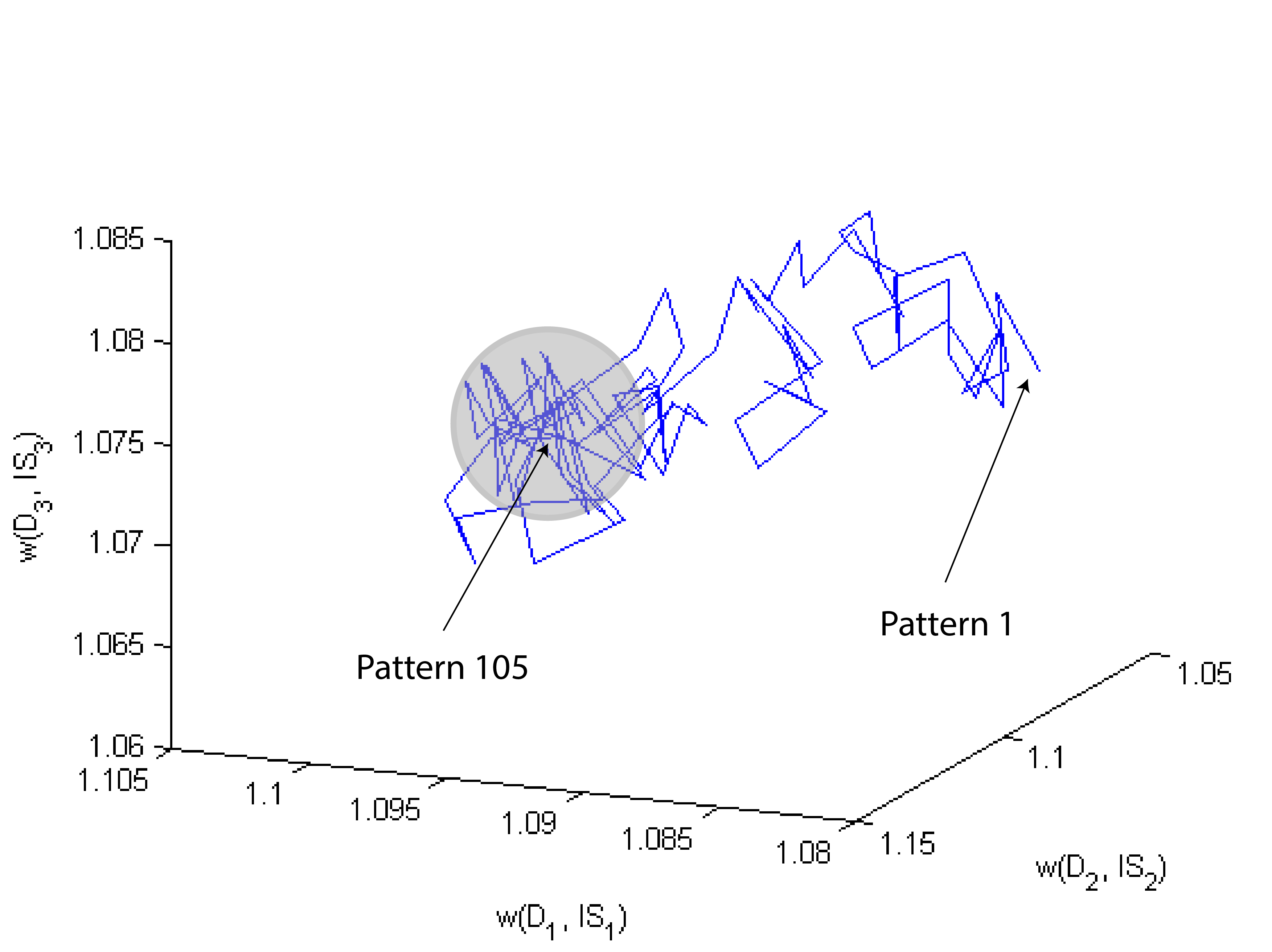}
\caption{Path of the weights along the presentation of the 105 patterns. A progressive stabilization of the weights is clearly noticeable (grey area) along the learning phase}
\label{FIG:CentroidPath}
\end{figure}

In the parameter initialization all weights were set to the same value. 
While new patterns were presented to the NMDS, the weights moved through a trajectory, depicted in Fig. \ref{FIG:CentroidPath}, achieving a progressive stabilization towards a combination of values that maximized the synchrony to the targets corresponding to the Go patterns.
In table \ref{tab:TableLabel} we report the results of the test performed on the trained MNSD.
\begin{table}[ht]
\caption{Test results}
\tiny
\centering % used for centering table
\begin{tabular}
{|| c|c|c|c|c|c|c|c|c ||}
\hline
Positive&Negative&True Positive&True Negative&False Positive&False Negative&Accuracy& Precision& Recall\\
(P)&(N)&(TP)&(TN)&(FP)&(FN)& & & \\
\cline{1-9}
23 & 22 & 16 & 14 & 8 & 7 & 67\% & 67\% & 70\%\\
\cline{1-9}
\end{tabular}
\label{tab:TableLabel}
\end{table}

We have considered the following formulas:
\begin{equation}
Accuracy= \frac {TP + TN}{TP + TN + FP + FN}
\end{equation}
\begin{equation}
Precision = \frac {TP}{TP + FP}
\end{equation}
\begin{equation}
Recall = \frac {TP}{TP + FN}
\end{equation}

\section{Discussion and Conclusions}

In this study we have presented a multineuronal spike-pattern detection structure, MNSD, which combines the LIFL neuron model and heterosynaptic STDP, to perform online learning and recognition of multineuronal spike patterns.\\
The presented structure includes a bio-plausible self-tuning mechanism, that is able to learn and recognize multineuronal spike sequences through repeated stimulation.
The time-amplitude conversion operated by the spike latency feature is one of the key operation principles of the structure, then the same task could not be performed by a simple LIF.
Heterosynaptic excitatory STDP is allowed by the lateral connections in the network.
It represents a mechanism to enhance synaptic transmission, or synapsis strengthening, and consequently the sensitivity to incoming sensory inputs \citep{Christie2006}. \\
To illustrate the ability of our structure, we have used the MNSD tool to discriminate between Go and NoGo decision during a motor-inhibitory task, obtaining good results. 
MNSD can be further applied to problems with a greater number of features, and in other contexts of temporal stream data where SNN have already been applied \citep{LoSciuto2016, Brusca2017}.\\
STDP is present in different areas of the brain, including sensory cortices like the visual and auditory, as well as the hippocampus \citep{Yu2013, Yu2014, Matsumoto2013}. 
Since STDP associates with coincidence detectors, where neurons get selective to a repetitive input pattern, it is thought to be crucial for memory and learning of the attributes of the stimuli (e.g., visual and auditory stimuli), even when the exposure is to meaningless sensory sequences that the subject is unaware of \citep{Masquelier2017}. 
Thus, the structure presented here may help understanding how humans learn repeating sequences in sensory systems. 
In fact, in sensory systems, different stimuli evoke different spike patterns but the exact way this information is extracted by neurons is yet to be clarified.\\
We can envisage to expand our MNSD structure in a modular way, such that each class is topologically structured with elementary building blocks among repetitive cortical columns and microcircuits: add other branches in parallel to increase the number of features, or inject the same $IS$ to more than one delay neuron to obtain articulated shapes of \emph{class hypervolumes}.\\

% Sviluppi futuri: Considerare sequenze di ``cardinalità di ramo'' > 1

\section{Supplementary material}

\subsection*{$T_{\lowercase{r}}$ calculation}

\begin{figure}

\centering
\includegraphics[width=0.7\textwidth]{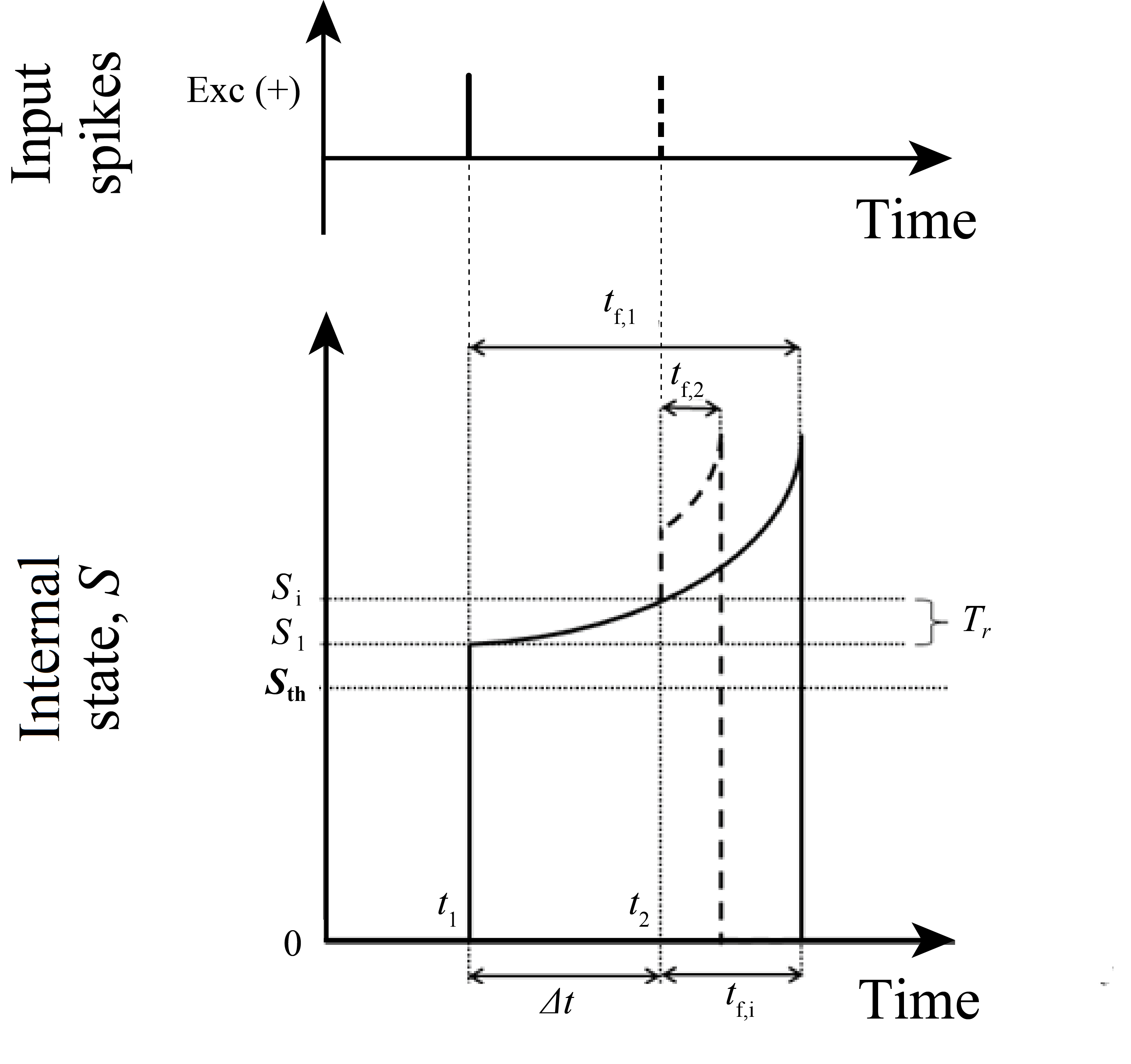}
 \caption{Representation of $T_r$. LIFL neuron in active mode is characterized by a spontaneous growth of $S$. If a pulse arrives before the actual spike generation, $S$ is modified and the $t_{f}$ will be recalculated. The recalculation considers the intermediate state $S_i$, i.e., the neuron state at the time the pulse arrives.} 
\label{Fig:S1_TrCalculation}
\end{figure}

Referring to Fig. \ref{Fig:S1_TrCalculation}, at the time the neurons inner state is altered from a second input (here excitatory, but non influencial to calculation purposes), the \emph{intermediate state} $S_i$ is determined, and then $T_{r}$ is calculated. 

Using the event-driven simulation technique the update of network elements happens only when they receive or emit a spike. Once an input spike arrives in active mode, the $S_i$ is calculated on the basis of the time remaining to the spike generation.

Referring to the generic inner state $S_{i}$ the firing equation is:

\begin{equation}\label{EQ:tfi}
t_{f,i}=\frac{1}{S_i-1}
\end{equation}

We define:

\begin{equation}
\Delta t=t_2-t_1
\end{equation}
where $t_1$ and $t_2$ represent the arrival instants of the synaptic pulses to the considered neuron. Then:

\begin{equation}\label{EQ:tfiNEW}
t_{f,i}=t_{f,1}-\Delta t 
\end{equation}

Rearranging Eq. \ref{EQ:tfi}, we obtain:
\begin{equation}\label{EQ:SiNEW}
S_i=\frac{1}{t_{f,i}} +1
\end{equation}

Now we combine Eq. \ref{EQ:tfiNEW} with Eq. \ref{EQ:SiNEW}
\begin{equation}\label{EQ:SiNEW2}
S_i=\frac{1}{t_{f,1}-\Delta t}+1
\end{equation}

By defining
\begin{equation}\label{EQ:Treq}
T_r = S_i - S_1
\end{equation}

where
\begin{equation}\label{EQ:S1}
S_1=\frac{1}{t_{f,1}} + 1
\end{equation}

and putting Eq .\ref{EQ:SiNEW2} and \ref{EQ:S1} in \ref{EQ:Treq}, we obtain:

\begin{equation}
T_r=\frac{1}{t_{f,1}-\Delta t}-\frac {1}{t_{f,1}}
\end{equation}

that can be rearranged as

\begin{equation}\label{EQ:DeltaSNEW}
T_r=\frac{\Delta t}{t_{f,1} (t_{f,1} -\Delta t) }
\end{equation}

Note that we are interested in determining an intermediate state; this implies that we consider the second synaptic pulse only if its timing (i.e., $t_2$) falls before the spike occurs. This gives us:
\begin{equation}\label{EQ:condition}
\Delta t < t_{f,1}
\end{equation}
thus we do not have restrictions from the denominator of \ref{EQ:DeltaSNEW}.

The relation \ref{EQ:DeltaSNEW} can be generalized to the case as more input modify the firing time; then, we can write
\begin{equation}
T_r =S_{ic}-S_{ip}=\frac{\Delta t_i}{t_{f,ip} (t_{f,ip} -\Delta t_i) }
\end{equation}

with

\begin{equation}
\Delta t_i=t_{ic}-t_{ip}
\end{equation}

where the subscript $ip$ stays for \emph{intermediate-previous} and $ic$ for \emph{intermediate-current}. 

We can also make explicit the dependence of $T_r$ from the previous state, by inverting $t_{f,ip}$ trough Eq. \ref{EQ:tfi}, obtaining:

\begin{equation}\label{EQ:DeltaSNEW_doubleintermediate}
T_r=\frac{(S_{ip}-1)^2 \Delta t}{1-(S_{ip}-1)\Delta t}
\end{equation}

Obviously, the same considerations on the arrival time of the second pulse remain valid, thus we do not have restrictions imposed by the denominator of \ref{EQ:DeltaSNEW_doubleintermediate}.

\subsection*{Conditions for the spike generation at the target neuron}

We hypothesize that contributions from delay neurons arrive ``sufficiently synchronous'' at the target (i.e., sufficiently synchronous to not allowing the internal state collapse to zero between two arrivals). Considering Fig.5 in the main text, the maximum value that is reached by the internal state of $T$ is $S_{M}$, that is given by:

\begin{equation}
\label{EQ:eqSM}
S_{M} = 3 - L_d \cdot \Delta t_{out,tot}
\end{equation}

where $\Delta t_{out,tot}$ is the time window between the first and the last spike reaching the target, evoked by a single sequence, equals to:

\begin{subequations}
\label{EQ:sist4}
\begin{equation}
\Delta t_{out,tot} = max (|\Delta t_{out_{D_1,D_2}}|,|\Delta t_{out_{D_1,D_3}}|) \; , \quad if \quad \Delta t_{out_{D_1,D_2}} \cdot \Delta t_{out_{D_1,D_3}} > 0
\end{equation}
\begin{equation}
\Delta t_{out,tot} = |\Delta t_{out_{D_1,D_2}}| + |\Delta t_{out_{D_1,D_3}}| \quad \quad \;,  \quad  if \quad \Delta t_{out_{D_1,D_2}} \cdot \Delta t_{out_{D_1,D_3}} < 0
\end{equation}
\end{subequations}

Now, considering the system of equations that relates the arrival times of the three contributions to T (see Eq.13 in the main text), we can explicitly write Eq.\ref{EQ:sist4} with respect to the input arrival times and weights afferent to $D_n$:

\begin{subequations}
\label{EQ:sist5}
\begin{align}
\Delta t_{out,tot} = & max (|\Delta t_{in_{D_1,D_2}}-\frac{1}{w(D_1, IS_1)-1}+\frac{1}{w(D_2, IS_2)-1}|, \nonumber \\
& |\Delta t_{in_{D_1,D_3}}-\frac{1}{w(D_1, IS_1)-1}+\frac{1}{w(D_3, IS_3)-1}|) \;  
\end{align}
\begin{align*}
\\ \quad if  \quad & (\Delta t_{in_{D_1,D_2}} - \frac{1}{w(D_1, IS_1)-1} + \frac{1}{w(D_2, IS_2)-1}) \cdot  (\Delta t_{in_{D_1,D_3}} - \nonumber \\
& \frac{1}{w(D_1, IS_1)-1} + \frac{1}{w(D_3, IS_3)-1}) > 0 \nonumber
\end{align*}
\begin{align}
\Delta t_{out,tot} = & |(\Delta t_{in_{D_1,D_2}}-\frac{1}{w(D_1, IS_1)-1}+\frac{1}{w(D_2, IS_2)-1}) - \nonumber \\ 
& (\Delta t_{in_{D_1,D_3}}-\frac{1}{w(D_1, IS_1)-1}+\frac{1}{w(D_3, IS_3)-1})| \quad \;  
\end{align}
\begin{align*}
\\ \quad if \quad & (\Delta t_{in_{D_1,D_2}} - \frac{1}{w(D_1, IS_1)-1} + \frac{1}{w(D_2, IS_2)-1}) \cdot (\Delta t_{in_{D_1,D_3}} - \nonumber \\ 
& \frac{1}{w(D_1, IS_1)-1} + \frac{1}{w(D_3, IS_3)-1}) < 0  \nonumber
\end{align*}
\end{subequations}

Then, given the specific input sequence, $S_M$ should result greater than $1+d$ in order to ensure that $T$ generates a spike. Considering Eq.\ref{EQ:eqSM}, we can then write the condition of activation of the target:

\begin{equation}
\label{EQ:cond7}
\Delta_{out,tot}<\frac{2-d}{L_d}
\end{equation}

In order to make Eq. \ref{EQ:cond7} explicit with respect to the input weights and spike intervals, we can state that if $\Delta t_{in_{D_1,D_2}}$ and $\Delta t_{in_{D_1,D_3}}$ have a concordant sign, the following relation ensures that $T$ spikes:

\begin{align}
& max (|\Delta t_{in_{D_1,D_2}}-\frac{1}{w(D_1, IS_1)-1}+\frac{1}{w(D_2,IS_2)-1}|,|\Delta t_{in_{D_1,D_3}}- \nonumber \\ 
& \frac{1}{w(D_1, IS_1)-1}+\frac{1}{w(D_3, IS_3)-1}|) <\frac{2-d}{L_d}
\end{align}

On the contrary, if $\Delta t_{in_{D_1,D_2}}$ and $\Delta t_{in_{D_1,D_3}}$ have discordant sign, then:

\begin{align}
& |(\Delta t_{in_{D_1,D_2}}-\frac{1}{w(D_1, IS_1)-1}+\frac{1}{w(D_2, IS_2)-1}) - (\Delta t_{in_{D_1,D_3}}- \nonumber \\ 
& \frac{1}{w(D_1, IS_1)-1}+\frac{1}{w(D_3, IS_3)-1})| <\frac{2-d}{L_d}
\end{align}

\section*{Conflict of Interest Statement}

The authors declare that the research was conducted in the absence of any commercial or financial relationships that could be construed as a potential conflict of interest.

\section*{Author Contributions}
G.S. designed the model and the computational framework; 
G.S., E.P., C.R.M. designed the experiment;
G.S., L.A.T., L.C., M.E.L., F.M., C.R.M. and E.P. wrote the paper;
F.M., L.A.T. provided and analyzed brain data;
M.E.L. contributed to shape the experiment.

\section*{Acknowledgments}
G.S. acknowledges financial support by the Spanish Ministry of Economy and Competitiveness (PTA-2015-10395-I).\\ 
Research by author L.C. is supported by Viera y Clavijo fellowship from Tenerife, Spain.\\
M.E.L. is supported by a postdoctoral fellowship from the Spanish Ministry of Economy and Competitiveness (IJCI-2016-30662).\\
C.R.M. and E.P. acknowledge support from the Spanish Ministerio de Econom\'{\i}a y Competitividad (MINECO) and Fondo Europeo de Desarrollo Regional (FEDER) through projects TEC2016-80063-C3-3-R (AEI/FEDER, UE).

\bibliographystyle{elsarticle-harv}

\end{document}